\documentclass[a4paper,fleqn,usenatbib]{mnras}

\usepackage[T1]{fontenc}
\usepackage{ae,aecompl}

\usepackage{graphicx}	
\usepackage{amsmath}	
\usepackage{amssymb}	
\usepackage{subfigure}
\usepackage{bm}

 \def\be{\begin{equation}}
\def\ee{\end{equation}}
\def\bea{\begin{eqnarray}}
\def\eea{\end{eqnarray}}

\newcommand{\J}{{\bm{\mathcal{J}}}}
\newcommand{\A}{{\bm{\mathcal{A}}}}

\newcommand{\R}{{\bm{\mathcal{R}}}}
\newcommand{\I}{{\bm{\mathcal{I}}}}
\renewcommand{\d}{{{\mathrm{d}}}}

\title[Accurately computing weak lensing convergence]{Accurately computing weak lensing convergence}

\author[S. M. Koksbang \& C. Clarkson]{
Sofie Marie Koksbang,$^{1}$\thanks{E-mail: sofie.koksbang@helsinki.fi}
Chris Clarkson,$^{2,3}$\thanks{E-mail: chris.clarkson@qmul.ac.uk}
\\
$^{1}$Department of Physics, University of Helsinki and Helsinki Institute of Physics, P.O. Box 64, FIN-00014 University of Helsinki, Finland\\
$^{2}$School of Physics \& Astronomy, Queen Mary, University of London, UK\\
$^{3}$Department of Mathematics and Applied Mathematics, University of Cape Town, Rondebosch 7701, South Africa
}

\date{Accepted XXX. Received YYY; in original form ZZZ}

\pubyear{2019}

\begin{document}
\label{firstpage}
\pagerange{\pageref{firstpage}--\pageref{lastpage}}
\maketitle

\begin{abstract}
Weak lensing will play an important role in future cosmological surveys, including e.g. Euclid and SKA. Sufficiently accurate theoretical predictions are important for correctly interpreting these surveys and hence for extracting correct cosmological parameter estimations. We quantify for the first time in a relativistic setting how many post-Born and lens-lens coupling corrections are required for sub-percent accuracy of the theoretical weak lensing convergence for $z\le 2$ (the primary weak lensing range for Euclid and SKA). We do this by ray-tracing through a fully relativistic exact solution of the Einstein Field Equations which consists of randomly packed mass-compensated underdensities of realistic amplitudes. We find that including lens-lens coupling terms and post-Born corrections up to second and third order respectively is sufficient for sub-percent accuracy of the convergence along $94\%$ of the studied light rays. We also find that a significant percentage of the studied rays have post-Born corrections of size over $10\%$ of the usual gravitational convergence, $\kappa^{(1)}$, and several rays even have post-Born corrections several times the size of $\kappa^{(1)}$ at $z = 2$.
\end{abstract}




\section{Introduction}
Effects of inhomogeneities on cosmological observations are described using cosmological perturbation theory. Traditionally, only the simplest part of the first order of these weak lensing expressions was considered but with the increasing precision of observational data, it has been necessary to go beyond first order perturbation theory. The first order perturbative lensing expressions are well known and full second order expressions for the convergence are given in e.g. \citet{key1,key2,Marozzi,Ido,Fanizza}. Third and fourth order corrections have only been considered partially, e.g. in \citet{iterative,krause}. However, there are several reasons why care must be taken when using perturbative expressions. First of all, the literature contains inconsistencies between analytical expressions obtained by different authors (see e.g. discussions in \citet{Marozzi,fanizza1,redshift}). Second of all, questions regarding the gauge dependence of lensing expressions must be carefully considered as pointed out in e.g. \citet{yoo1}. Lastly,  higher order corrections can be more important than lower order corrections because the real universe is better described by a weak field approximation than standard perturbation theory (see e.g. \citet{gevolution}). This obscures the validity of precision estimates based on perturbation theory truncated at a finite order. These points emphasize the importance of checking the appropriateness of perturbative expressions by a comparison with exact results. The straightforward way of doing this is to compare approximate results based on perturbation theory with exact results obtained by ray tracing through exact solutions to Einstein's equations. This permits one to assess whether or not the studied expressions are correct and yield the expected precision. This is the goal of the present work which is based on considering a semi-realistic inhomogeneous but statistically homogeneous and isotropic dust+$\Lambda$ solution to Einstein's equations. The model is constructed as a Swiss-cheese model. Swiss-cheese models are obtained by removing patches in an FLRW model (the ``cheese") and replacing them by inhomogeneous solutions to Einstein's equations. As Swiss-cheese models are exact solutions to Einstein's equations their metrics are known and hence light propagation can be described by exact means in these models. By choosing different metrics for constructing the Swiss-cheese model, different scenarios can be studied. For instance, Kottler structures have been glued together with $\Lambda$CDM models to study how observations based on thin beams are affected by small-scale inhomogeneities \citet{fleury1, fleury2}. Here, we are concerned with weak lensing and hence the Swiss-cheese models must be constructed to model the large-scale structures of the Universe. An appropriate Swiss-cheese model can be obtained by combining the $\Lambda$CDM model with Lemaitre-Tolmann-Bondi (LTB) voids. The LTB structures are distributed randomly so that the obtained results do not contain artifacts that are known to plague Swiss-cheese models with regularly distributed structures \citet{Vanderveld}. The model is used to identify the terms necessary to achieve perturbative results deviating less than $\sim 1 \%$ from the exact convergence, $\kappa$, up to a redshift of $2$ (the primary weak lensing range for several upcoming surveys goes out to $z \approx 2$). The results indicate that one must include lens-lens coupling at second order and post-Born corrections at second and third order to achieve a $~1\%$ precision out to $z = 2$. These corrections are derived following procedures similar to those of \citet{iterative} but are given in a form that generalizes the expressions to arbitrary order. Higher order corrections are also needed for sub-percent accuracy of the weak lensing shear but as the shear can be obtained from the convergence (see e.g. \citet{shear_review}) we do not investigate the necessary higher order corrections to the shear here.

\section{Light propagation}
The starting point for gravitational lensing is the geodesic deviation equation for a deviation vector $\xi^a$ which separates neighboring null geodesics on a future pointing bundle $k^a$: $\ddot\xi^a+k^b\xi^ck^d{R^a}_{bcd}=0$. We use $R^a_{~bcd}$ to denote the Riemann tensor, and dots indicate derivatives with respect to the affine parameter $\lambda$ associated with $k^a$. Given an observer, the equation can be projected into a 2d screen space spanned by an orthonormal tetrad basis $e_a^A$, $A=1,2$. The equation then becomes $\ddot\xi_A=\mathcal{R}_{AB}\xi^B$, where the optical tidal matrix is given by $\mathcal{R}_{AB}=  -\frac{1}{2}\delta_{AB} R_{cd}k^ck^d-C_{AcBd}k^ck^d$, with $R_{ab}$ the Ricci tensor and $C^a_{~bcd}$ the Weyl tensor. Solutions are written in terms of the Jacobi map, $\mathcal{J}_{AB}$, as $\xi_A(\lambda)=\mathcal{J}_{AB}(\lambda)\zeta^B$
where $\mathcal{J}_{AB}$ satisfies $\ddot{\mathcal{J}}_{AB}=\mathcal{R}_{AC}\mathcal{J}^C_{~B}$, with $\mathcal{J}_{AB}(\lambda_o)=0$ and $\dot{\mathcal{J}}_{AB}(\lambda_o)=-\delta_{AB}$. The differential equation can also be written as an integral equation which in matrix notation becomes $\J=(\lambda_o-\lambda_s)\I+ \int_{\lambda_o}^{\lambda_s}\d\lambda(\lambda_s-\lambda)\R\cdot\J$.
The Jacobi map takes the observed angle between two rays at the observer, $\zeta^A:=-\dot\xi^A|_\text{o}$, and maps it to the deviation vector at the source. The map can be expanded in terms of a background distance $\bar d_A$ and an amplification matrix $\mathcal{A}_{AB}=(1-\kappa)\delta_{AB}+\gamma_{AB}$ or (ignoring a small anti-symmetric part), 
\be
\J = \bar d_A \A
=\bar d_A\left(\begin{array}{cc}1-\kappa-\gamma_1 & -\gamma_2 \\-\gamma_2 & 1-\kappa+\gamma_1\end{array}\right).
\ee 
The convergence is therefore $\kappa = 1-\mathrm{tr}\,\J/2\bar d_A  $, and the shear is the trace-free part of $\A$. The angular diameter distance as a function of affine parameter is $d_A=\sqrt{\det \J} = \bar d_A\sqrt{(1-\kappa)^2-\gamma^2}$, where a bar denotes background quantities and $\gamma^2 :=\gamma_1^2+\gamma_2^2$.

We shall now consider solving for the convergence perturbatively, assuming a weak field approximation. 
We consider a perturbed FLRW metric (using $c = 1 = 4\pi G $): $\d  s^2 =
 a^2\big[-(1 + 2\Phi )\d \eta^2 
 + (1-2 \Psi)\gamma_{i j}\d x^{i}\d x^{j}\big]$,
where $\gamma_{ij} = \delta_{ij}$ if $x^i$ are Cartesian ($i,j,k,\ldots$ denote spatial indices). As only the sum of the potentials will appear we introduce the Weyl potential $\psi := (\Phi+\Psi)/2$. We will work on the conformal geometry, ignoring factors of $a$ except where they appear in relation to $\delta\rho$. In this case the background angular diameter distance is $\bar d_A=r$, and we will use $r$ as the affine parameter distance down the past light cone from observer to source: $r=\lambda_o-\lambda_s$. 
For the considered metric, the leading order perturbative contribution to the optical tidal matrix is
\be
\delta \mathcal{R}_{AB} = -\delta_{AB}\delta\rho/a^2-2\nabla_{\langle A}\nabla_{B\rangle}\psi\,.
\ee
Note that we keep only the terms with the highest number of derivatives in screen space since these terms are by far the most important for normal lensing events. Angle brackets mean the trace-free part of a tensor: $X_{\langle AB\rangle}=X_{AB}-\frac{1}{2}\delta_{AB} \delta^{CD}X_{CD}$, and $\nabla^2=\nabla_A\nabla^A$. (As we keep only highest derivatives of the potential, no derivatives of the tetrad appear.) If we identify $\psi$ as the full gravitational potential (e.g., from an N-body simulation) rather than the potential from linear perturbation theory, this is an excellent approximation. We use the flat sky approximation so the derivatives in the above expression can simply be swapped with Cartesian derivatives in the final expressions. Note lastly that the Ricci term deviates from the exact Ricci term only in terms of Born corrections and Doppler corrections as the exact Ricci term is $-\frac{1}{2}\delta_{AB} R_{cd}k^ck^d\propto -\rho \left( k^{\mu}u_{\mu}\right) ^2$. 

Using $\delta \mathcal{R} $, the integral equation for the amplification matrix becomes
\be
\A=\I+ \int_{0}^{r}\d r'\frac{r'(r-r')}{r}\delta \R\cdot\A\,.
\ee
Our goal is to solve this equation to obtain $\kappa$ at sub-percent accuracy as will be necessary for the correct interpretation of and forecasts for upcoming surveys. We define the operator 
\be
\mathcal{O}_A^{~B}(r,r')=-\int_{0}^{r}\d r'\frac{r'(r-r')}{r}\delta \mathcal{R}_A^{~B}(r')\,.
\ee
Repeated substitution then shows that the solution can be written as the series
\bea
\mathcal{A}_{A}^{~B}(r) &=& \delta_{A}^{~B}-\mathcal{O}_{A}^{~B}(r,r') + \mathcal{O}_{A}^{~C}(r,r')\mathcal{O}_{C}^{~B}(r',r'')-\nonumber\\&&  
\mathcal{O}_{A}^{~C}(r,r')\mathcal{O}_{C}^{~D}(r',r'')\mathcal{O}_{D}^{~B}(r'',r''')+\cdots\,.
\eea
The second term gives the standard linear convergence and shear:
\bea
\kappa(r) &=& \int_{0}^{r}dr'\frac{r'(r-r')}{r}a^2\delta\rho(r')\,,\\
\gamma_{AB}(r)&=& 2\int_{0}^{r}dr'\frac{r'(r-r')}{r}\nabla_{\langle A}\nabla_{B\rangle}\psi(r')\,,
\eea
with $-\gamma_{22}=\gamma_{11} = \gamma_1$, $\gamma_{12}=\gamma_{21} = -\gamma_2$.

Promoting these to operators, $\kappa(r)\to\kappa(r,r')$ and $\gamma_{AB}(r)\to\gamma_{AB}(r,r')$, we can separate the trace and trace-free parts of $\mathcal{O}_{AB}$ as
\be
\mathcal{O}_{AB}(r,r') = \kappa(r,r')\delta_{AB}+\gamma_{AB}(r,r')\,. 
\ee
Using this we can now calculate the higher order contributions to the convergence and shear by extracting the trace and trace-free parts of the higher-order products of $\mathcal{O}_{AB}$. In general, a contraction of 2 symmetric matrix operators in 2d, $X_{AB}=X\delta_{AB}+\hat X_{AB}$, can be expanded into its trace and trace-free parts as
\bea
X_{A}^{~C}Y_{CB} &=& \frac{1}{2}(2XY+\hat X^{CD}\hat Y_{CD})\delta_{AB}+ 
\nonumber\\&&
X\hat Y_{AB}+\hat X_{AB}Y+\hat X_{\langle A}^{~~C}\hat Y_{B\rangle C}\,.
\label{jdsnckscns}
\eea
We can use this repeatedly to calculate all the higher-order contributions we require. 
Consequently we have the second-order convergence in terms of the first-order operators $\kappa$ and $\gamma_{AB}$:
\bea
&\kappa^{(2)}(r)& = \kappa_{\kappa\kappa}+\kappa_{\gamma\gamma} \nonumber\\
&=& -\kappa(r,r')\kappa(r',r'')-\frac{1}{2}\gamma^{AB}(r,r')\gamma_{AB}(r',r'')
\nonumber\\
&=&
- \int_{0}^{r}dr'\frac{r-r'}{r}a^2\delta\rho(r')\int_{0}^{r'}dr''\left( r'-r''\right) r''a^2\delta\rho(r'')\nonumber\\
&-& 2\int_{0}^{r}dr'\frac{r-r'}{r}\nabla^{\langle A}\nabla^{B\rangle}\psi(r')
\times
\nonumber\\
&&\int_{0}^{r'}dr''\left( r'-r''\right) r''\nabla_{\langle A}\nabla_{B\rangle}\psi(r'').
\eea
Similarly we can read off the correction to the shear from the trace-free part as
\bea
\gamma^{(2)}_{AB}&=&\kappa(r,r')\gamma_{AB}(r',r'')
+\gamma_{AB}(r,r')\kappa(r',r'')
+\nonumber\\
&&\gamma_{\langle A}^{~~C}(r,r')\gamma_{B\rangle C}(r',r'')
\nonumber\\
&=& 4\int_{0}^{r}dr'\frac{r-r'}{r}\nabla_{\langle A}\nabla^{C}\psi(r')\times
\nonumber\\
&&\int_{0}^{r'}dr''\left( r'-r''\right) r''\nabla_{ B\rangle}\nabla_{C}\psi(r'').
\eea
For the third-order terms we now use $\mathcal{O}_{A}^{~C}(r,r')\mathcal{O}_{C}^{~D}(r',r'') = \kappa^{(2)}(r,r',r'')\delta_{AB}+\gamma^{(2)}_{AB}(r,r',r'')$,
again promoting $\kappa^{(2)}$ and $\gamma_{AB}^{(2)}$ to operators and using~\eqref{jdsnckscns}:
\bea
\kappa^{(3)}(r)=\kappa^{(2)}(r,r',r'')\kappa(r'',r''')+\nonumber\\
\frac{1}{2}\gamma^{(2)}_{AB}(r,r',r'')\gamma^{AB}(r'',r''')\nonumber\\
=-\kappa(r,r')\kappa(r',r'')\kappa(r'',r''')-\nonumber\\
\frac{1}{2}\gamma^{AB}(r,r')\gamma_{AB}(r',r'')\kappa(r'',r''')\nonumber\\
+\frac{1}{2}\kappa(r,r')\gamma_{AB}(r',r'')\gamma^{AB}(r'',r''')+\nonumber\\
\frac{1}{2}\gamma_{AB}(r,r')\kappa(r',r'')\gamma^{AB}(r'',r''')+\nonumber\\
\frac{1}{2}\gamma_{\langle A}^{~~C}(r,r')\gamma_{B\rangle C}(r',r'')\gamma^{AB}(r'',r''')\nonumber\\
\eea
And so on. In this way we can explicitly see the different contributions to the higher-order convergence.
The third order lens-lens coupling correction of the convergence is not necessary for percent accuracy at $z<2$ in the studied model but could be necessary at higher redshifts.

The integrals in the above expressions are in principle to be taken along the real perturbed line of sight, but this makes them difficult to compute. Alternatively, we can simply interpret these integrals as being along the background light ray, an approximation known as the Born approximation. We can then obtain corrections to the Born approximation by expanding around the background line of sight as follows.

It follows from the geodesic equation that deviations, $\delta x^A$, to the light path in screen space are given by $\delta x^A = -2\int_0^rdr'(r-r')\psi_{,A}(r')$. Different orders of $\delta x^A$ are obtained by Taylor expanding $\psi_{,A}$ around the unperturbed light path under the integral, while integrating according to the Born approximation, i.e. for the first and second order:
\bea
\delta x^{(1)A}
&= & -2\int_0^rdr'(r-r')\psi_{,A}(r')\nonumber\\
\delta x^{(2)A}
&= & -2\int_0^rdr'(r-r')\psi_{,AB}(r')\delta x^{(1)B}(r') \nonumber\\
&= & 4\int_0^rdr'(r-r')\psi_{,AB}(r')\int_0^{r'}dr''(r'-r'')\psi^{,B}(r'').\nonumber\\
\eea
Inserting these into the lowest perturbative expression for the amplification matrix gives the following corrections to the convergence:
\bea
\kappa_{B1}  
&= & -2\int_{0}^{r}dr'\frac{r-r'}{r}r'a^2\delta\rho_{,A}\int_{0}^{r'}dr''(r'-r'')\psi^{,A}\nonumber\\
\kappa_{B2} \\
&= & 2\int_{0}^{r}dr'\frac{r-r'}{r}r'a^2\delta\rho_{,AB}\times\nonumber\\
&&\int_{0}^{r'}dr''(r'-r'')\psi_{,A}\int_{0}^{r'}dr''(r'-r'')\psi_{,B} \nonumber\\
&+ & 4\int_{0}^{r}dr'\frac{r-r'}{r}r'a^2\delta\rho_{,A}\times\nonumber\\
&&\int_{0}^{r'}dr''(r'-r'')\psi_{,AB}
\int_{0}^{r''}dr'''(r''-r''')\psi_{,B}\nonumber\\
\eea

\begin{figure}
\includegraphics[scale = 0.9]{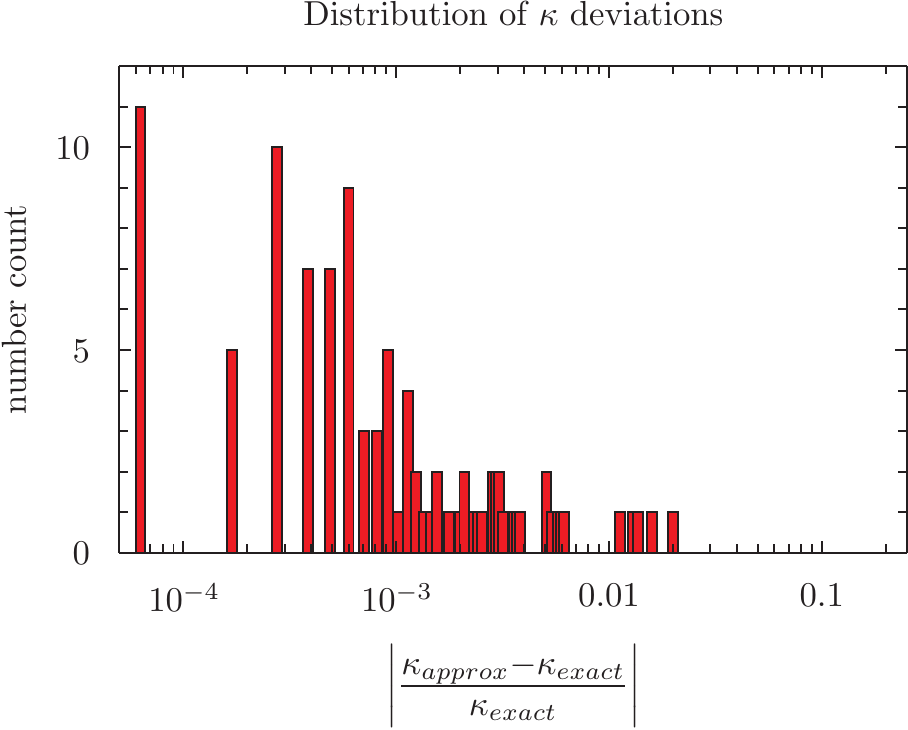}
\caption{Precision of $\kappa_{approx}:=\kappa^{(1)}+\kappa_v+\kappa_{B1}+\kappa_{B2}+\kappa_{\kappa\kappa}+\kappa_{\gamma\gamma}$ along 100 light rays at $z=2$. $\kappa_{exact}$ is the exact convergence computed using exact light propagation with the exact Swiss-cheese spacetime.}
\label{fig:ddahist}
\end{figure}

Lastly, it is necessary also to include the Doppler convergence $\kappa_{v}$ in the computations (see e.g. \citet{BrightSide}). The lowest order Doppler convergence is given by
\begin{equation}
\kappa_v = \left( 1 - \frac{1}{ra_{,t}}\right)\left( \mathbf{v}_S-\mathbf{v}_O\right)\cdot \mathbf{n} +\mathbf{v}_O\cdot\mathbf{ n},
\end{equation}
where $\mathbf{n}$ is the direction vector of the light ray computed in the background and $\mathbf{v}_S$ and $\mathbf{v}_O$ are the spatial velocity fields of the source and observer, respectively.

\begin{figure}
\includegraphics[scale = 0.9]{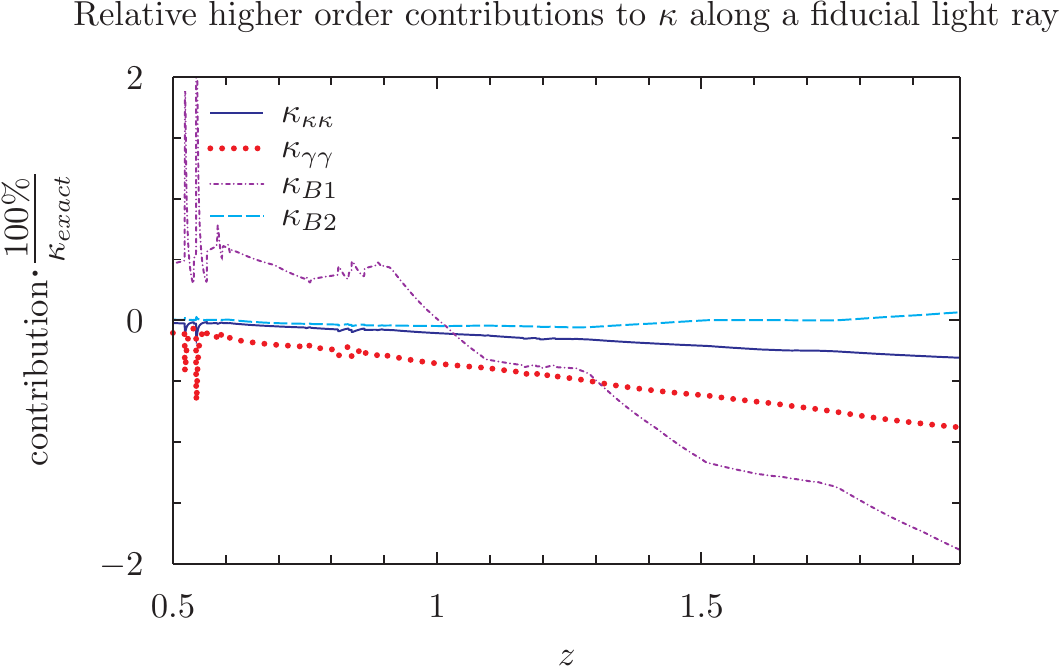}
\caption{Higher order contributions to $\kappa$ along a fiducial light ray in the Swiss-cheese model. The contributions are shown as percentages of the exact convergence $\kappa_{exact}$. The figure does not show the area with $z< 0.5$ as this area is obscured by peaks due to division by zero and mismatches of the exact and approximate Doppler convergences. The mismatches in the Doppler convergence will not be discussed here where accumulative effects along the light rays are the focus.}
\label{fig:lr4}
\end{figure}

\section{Results}
We show the above discussed contributions to the convergence along light rays propagated in a Swiss-cheese model based on the spherically symmetric Lemaitre-Tolman-Bondi (LTB) dust solutions. The LTB structures are specified by a $\Lambda$CDM background (the ``cheese") with $\Omega_{\Lambda} = 0.7$ and $H_0 = 70$km/s/Mpc, a constant big bang time and a curvature parameter given by
\newline
\begin{equation}
k(r) = \left\{ \begin{array}{rl}
-5.57\cdot 10^{-8}r^2\left(\left(\frac{r}{40\text{Mpc}} \right)^6 -1 \right)^6  &\text{if} \,\, r<40\text{Mpc} \\
0 &\mbox{ otherwise}
\end{array} \right..
\end{equation}
The choice of $k$ specifies the shape and size of the LTB structures. While the detailed shape in terms of e.g. steepness of the resulting density contrast is not particularly important for the results, $k$ must be chosen to yield structures of overall qualities similar to realistic large-scale structures. The choices specified above yields structures of semi-realistic sizes: Voids with radius $\sim 30$Mpc and minimum density $\delta\rho/\rho\approx0.3$ surrounded by a mass compensating shell with overdensity peaking at $\delta\rho/\rho\sim 100$. The LTB structures are arranged randomly according to the description in \citet{koksbangCMB}, leading to a Swiss-cheese universe of random close-packed LTB structures with a packing fraction of $\sim 0.6$. The exact convergence along light rays in the Swiss-cheese model are compared with the perturbative approximations obtained from mock N-body reproductions of the Swiss-cheese model following the mapping procedure of \citet{koksbangMap1,koksbangMap2}. Within that setting, light rays can be traced using standard perturbation theory on a flat FLRW background combined with the non-linear density and velocity fields of the mock N-body data.

\begin{figure}
\includegraphics[scale = 0.9]{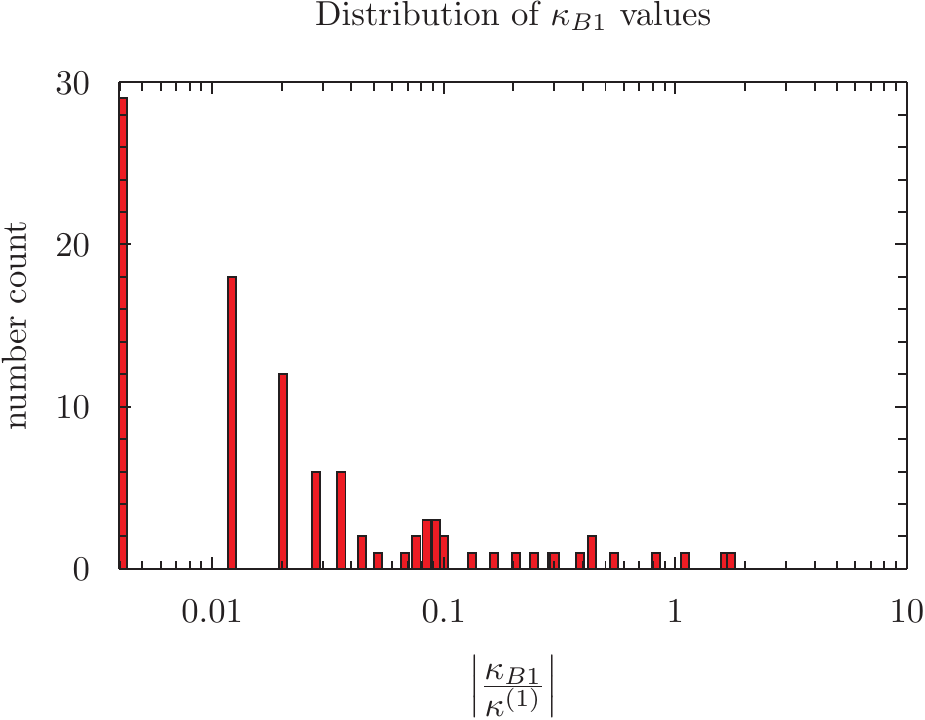}
\caption{Size of $\kappa_{B1}$ relative to $\kappa^{(1)}$ at $z = 2$.}
\label{fig:bornhist}
\end{figure}
100 random light rays have been considered, each with an observer in the $\Lambda$CDM background. By including the lowest order lens-lens coupling and the two lowest orders of post-Born corrections as well as $\kappa^{(1)}+\kappa_{v}$, the difference between the exact and approximate convergence reduces to less than $1\%$ in the studied redshift interval along 94 of the light rays when local effects from the Doppler convergence are ignored. This is illustrated in figure (\ref{fig:ddahist}) which shows the deviation between the exact and approximate $\kappa$ at $z = 2$ (since the deviations between exact and approximate results generally increase with redshift, showing the results at $z = 2$ is appropriate). In general, we find that each of the higher order corrections individually reach $1\%$ of both $\kappa^{(1)}$ and $\kappa_{exact}$ along at least one of the studied light rays. An example light ray is given in figure \ref{fig:lr4} which shows the individual contributions to the convergence for one of the light rays. For this particular line of sight, the second and third order contributions are all of similar size and are $\sim 1\%$ or sub-percent individually while their sum is approximately $3.5\%$ of the exact convergence. As indicated already in this figure, the lowest order post-Born correction is by far the most important contribution after $\kappa^{(1)}+\kappa_v$. The distribution of the lowest order Born correction along all the light rays at $z = 2$ is shown in figure \ref{fig:bornhist}. Quite noticeably this second order contribution, which is often neglected, exceeds $10\%$ of the traditional gravitational convergence, $\kappa^{(1)}$, along multiple rays and it even exceeds the value of $\kappa^{(1)}$ along several light rays. This supports the findings of e.g. \citet{fanizza2} that the lowest order Born correction leads to significant effects on at least some observations.

\section{Conclusions}
By comparing the approximate weak lensing convergence with its exact counterpart we showed that a sub-percent accuracy up to $z \approx 2$ of the weak lensing convergence requires several orders of the post-Born correction as well as the lowest order lens-lens coupling correction. Since most contributions to the convergence are in the form of integrals along the line of sight, even higher order corrections are needed if the same accuracy is required at higher redshifts. Our results thus indicate that the correct treatment of upcoming surveys will require the inclusion of several higher order corrections to the standard gravitational convergence $\kappa^{(1)}$. Our results also indicate that already for current surveys, including the post-Born correction at lowest order is important as it can become the dominant contribution to the convergence along some lines of sight.
\newline\indent
The presented results were obtained for a specific Swiss-cheese model based on a single, specific LTB structure that reduces exactly to the background $\Lambda$CDM model at its edge. Such a set-up is clearly overly simplified compared to the real universe but the LTB structures were distributed {\em randomly} and close-packed with a structure size and amplitude chosen to approximate realistic large-scale structures. Hence, while there would presumably be small adjustments in the results obtained here if using another (semi-)realistic model, the overall results are expected to be valid in general for a universe with a soap-bubble large-scale structure distribution with an average $\Lambda$CDM evolution. None the less, it would be interesting to further investigate the results obtained here e.g. by using the method described here for computing higher order corrections along light rays in N-body simulations. Any significant quantitative discrepancies with the results obtained here would require an explanation while an overall agreement with the results presented here would strengthen the credibility of results obtained by ray tracing through N-body simulations. In relation to this it would also be useful to use the method studied here for computing $\kappa$ to evaluate the accuracy of the multiple-lens approximation often used for ray tracing through N-body simulations.

\section*{Acknowledgements}
We thank Thomas Tram and Jeppe Dakin for correspondence on FFTW3 and Giovanni Marozzi for correspondence on the work presented in \citet{Marozzi}. We also thank the anonymous referees for useful comments that have significantly improved the manuscript.
\newline\indent
SMK is supported by the Independent Research Fund Denmark under grant number 7027-00019B. CC was supported by STFC Consolidated Grant ST/P000592/1.
\newline\indent
The work has been done using computer resources from the Finnish Grid and Cloud Infrastructure urn:nbn:fi:research-infras-2016072533.












\bsp	
\label{lastpage}
\end{document}